\documentclass[twocolumn,showpacs,preprintnumbers,amsmath,amssymb]{revtex4-1}
\usepackage{graphicx}
\usepackage{dcolumn}
\usepackage{bm}
\usepackage{color}
\usepackage{amsmath, amssymb, gensymb}
\usepackage{subfigure}
\usepackage[svgnames*,table,dvipsnames]{xcolor} 
\usepackage{ulem}

\definecolor{RED}{rgb}{1,0,0}

\newcommand{\JB}[1]{\textcolor{black}{#1}}
\newcommand{\JBn}[1]{\textcolor{red}{#1}}

\begin{document}
\preprint{APS/123-QED}

\title{Soft grain compression: beyond the jamming point}

\author{Thi-Lo Vu}
\author{Jonathan Bar\'{e}s} \email{jonathan.bares@umontpellier.fr}

\affiliation{Laboratoire de M\'{e}canique et G\'{e}nie Civil, Universit\'{e} de Montpellier, CNRS, Montpellier, France}

\begin{abstract}

We present the experimental studies of highly strained soft bidisperse granular systems made of hyperelastic and plastic particles. We explore the behavior of granular matter deep in the jammed state from local field measurement\JB{,} from the grain scale to the global scale. By mean of \JB{a dedicated digital image correlation code and an accurate image recording method,} we measure for each compression step the evolution of the particle geometries and their right Cauchy-Green strain tensor fields. We analyze the evolution of the usual macroscopic observables (stress, packing fraction, coordination, fraction of non-rattlers, \textit{etc}.) along the compression process through the jamming point and far beyond. \JB{Analyzing the evolution of the local strain statistics, we evidence} a crossover in the material behavior deep in the jammed state \JB{for both sorts of particles}. We show that this crossover \JB{is due to a competition between material compression, dilation and shear, so its position} depends on the particle material. We argue that the strain field is a reliable observable to describe the evolution of a granular system through the jamming transition and deep in the dense packing state whatever is the material behavior. 

\end{abstract}

\maketitle

\section{Introduction} \label{intro}

Wound healing \cite{martin2004_dev}, cell monolayer growth \cite{angelini2011_pnas,madhikar2018_arx}, embryo developing \cite{atia2017_arx,sadati2013_dif}, cancer invasion \citep{bi2015_nat,bi2016_prx}, squeezed foam \cite{bolton1990_prl,durian1995_prl,katgert2010_epl,winkelmann2017_csa}, \JB{squeezed} emulsion \cite{brujic2003_fd,brujic2003_pa,zhou2006_sci,brujic2007_prl,desmond2013_sml,lorenzo2013_fh,fan2017_sml} and metal, plastic or ceramic powder sintering \cite{bares2014_prl}; all these very different systems have the common point to be described as packings of discrete, deformable particles. Even if the jamming transition -- the transition between a fluid-like \JB{and} a solid-like behavior of the granular matter -- of stiff \JB{or weakly deformable} particles has been very well studied, theoretically \cite{liu1998_nat,degiuli2015_jcp,jacquin2011_prl,berthier2011_pre}, numerically \cite{ohern2003pre,ohern2002_prl,heussinger2009_prl,srivastava2017_sm,dagois2017_sm,kapfer2015_prl} and experimentally \cite{brujic2007_prl,bi2011_nat,majmudar2007_prl,behringer2015_crp,zhao2019_arx} during the past decades, very few is known about the behavior of these soft granular systems \JB{at} high packing fraction.

Up to the jamming point the compressibility of these particles can be neglected since the material is not able to carry load. Around the jamming point it is possible to consider that these particles are ruled by a linear elastic behavior since the \JB{stress and so the deformations are assumed to be small} \cite{brujic2003_fd,brujic2003_pa,zhou2006_sci}. However far beyond the jamming point it is necessary to consider a more complicated local behavior for the particle materials; different mechanisms more complicated than elasticity are at play: hyperelasticity, plasticity, damage, fracture \textit{etc}. \JB{This implies that a large diversity of mechanical behaviors have to be considered for the particles. On top of this,} it is also important to consider \JB{a large} diversity of scales, shapes and interaction mechanisms. 

These soft granular matter systems have been studied experimentally and numerically in the biologic context to understand the effect of mechanics on a the cell arrest, deformation and interaction near the jamming point. This aims at understanding the embryonic development, cancer invasion, and wound healing \cite{martin2004_dev,sadati2013_dif,angelini2011_pnas,szabo2006_pre}. Experiments have also been carried out to make faceted polyhedral microgels objects from emulsions compressed to packing fractions very close to $1$ \cite{fan2017_sml}. But monolayer soap foams \JB{and oil emulsion} are most likely the most well understood soft granular systems since many studies have been carried out near the jamming transition showing that, at the global \JB{and local scales, these systems do not differ too much from rigid particle packings in terms of coordination and force distributions \cite{brujic2003_pa,bolton1990_prl,durian1995_prl,brujic2007_prl,zhou2006_sci,katgert2010_epl,winkelmann2017_csa}}. However, few is known for bulk material particles and not so many methods exist to measure mechanical interactions at the grain scale. Among differents\cite{howell1999_prl,jongchansitto2014_sm,brodu2015_nat,amon2017_aip} the most common is the \JB{photoelasticimetry \cite{howell1999_prl,daniels2017_rsi,all2019_wiki,zadeh2019_arx}, the measurement of the boundary deformation \cite{brujic2003_fd,brujic2003_pa,brujic2007_prl,brodu2015_nat} and recently inverse problem method coupled with digital image correlation (DIC) have been also used for stiff particles \cite{hurley2014_jmps,marteau2017_gm,li2019_gm}. However, nothing exists for local (sub-granular) measurements in soft granular systems in an highly strained state}. 

Recently numerical models have manage to simulate these squeezable materials. Several approaches have been tested: \citet{nezamabadi2017_gm} have introduced the Material Point Method \JB{(so called MPM)} and the Bonded Particle Model \JB{(so called BPM)} to simulate the compression of dense granular systems. They have been followed by \citet{boromand2018_prl} who introduced the Deformable Particle Model \JB{(so called DPM)} to simulate the compression of 2D grains which are polygons composed of a large number of vertices. \JB{Nevertheless, these different studies have barely been confronted to experimental results \cite{vu2019_pre}} and they all only focus on particle geometry variations and macroscopic observable changes \JB{without looking at the material evolution down to the particle scale}. 

However several questions remain about the behavior of highly strained granular matter: ($i$) Is there a transition or a crossover in the material behavior when the packing fraction increase in the dense state? ($ii$) Are some elements of the Hertz contact framework still valid in this state? ($iii$) What is the effect of the exact nature of the material whose the particles are made of, on the local and global behaviors of the packing? In order to tackle these issues, in this paper, we propose a pioneer model experiment bringing informations currently inaccessible. We introduce a novel method capable of measuring the local displacement field and inferring other mechanical fields in irregularly-shaped highly strained granular materials of any rheology. These measurements are made at the grain scale without inverse problem solving, so without any assumption on the system fabric, \JB{nor on the material rheology}. 

The paper is organized as follow. First, we introduce the experimental method and image processing techniques. Then, we present the results in a second \JB{section} giving first the evolution of the global observables and then the variation of the local statistics. Finally we end with concluding discussions in a third \JB{section}.

\section{Experimental method} \label{expeMeth}

\subsection{Set-up} \label{expe}

\begin{figure}[htb!]
	\begin{center}
		\includegraphics[width=0.49\textwidth]{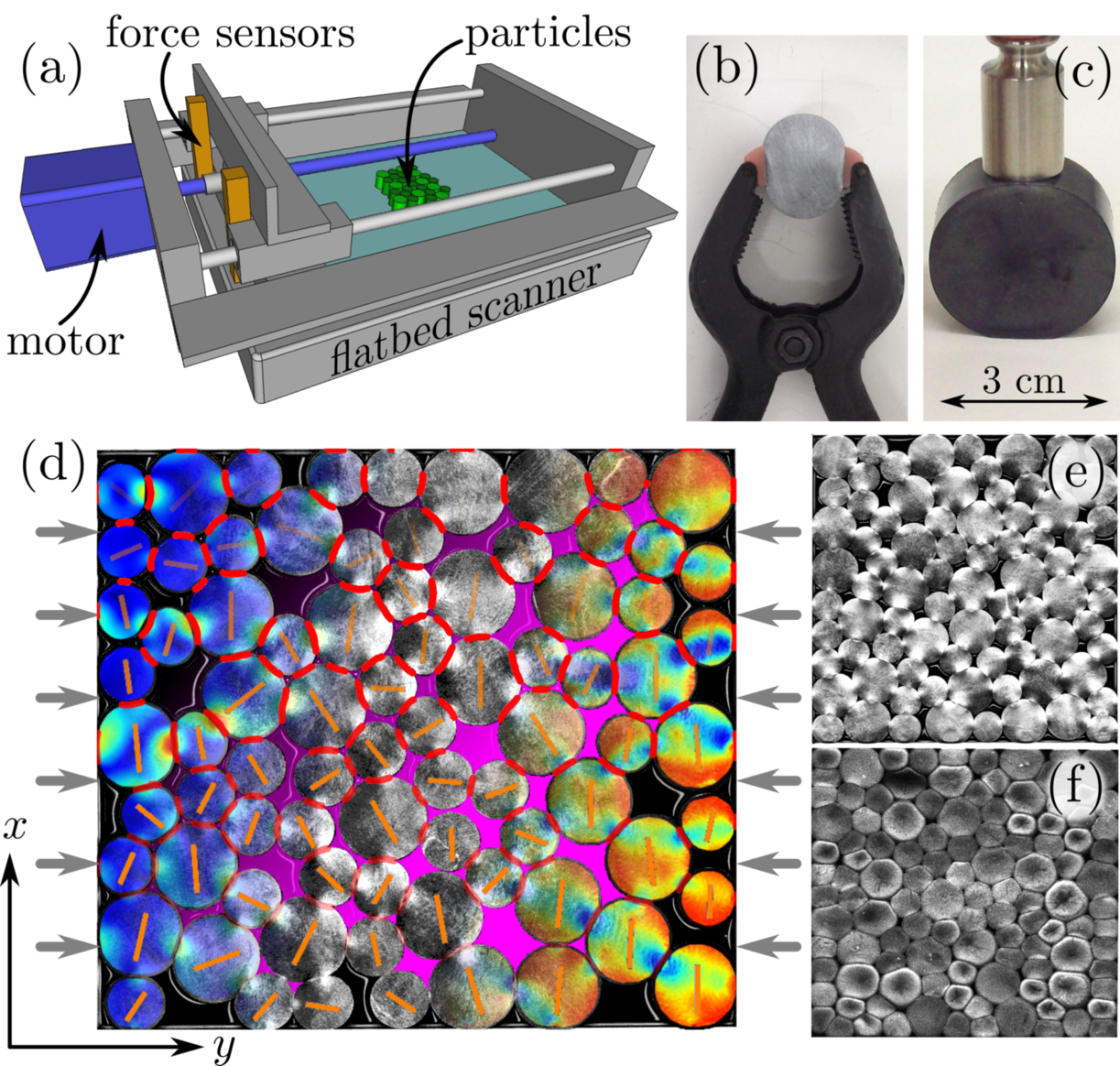}
	\end{center}
\caption{(Color online) a: Schematic view of the experimental set-up. A collection of bidisperse soft grains is compressed stepwise on a flatbed scanner recording the applied force and imaging the packing evolution down to a very small scale. b-c: Pictures of a hyperelastic (b) and a plastic (c) particles made of silicone \cite{silicone} and agar gel \cite{agar} compressed with force $20$~N and $0.2$~N respectively. d: Compositive view of compressed packing of hyperelastic particles: magnitude displacement field (left), raw scanned image (middle) and $\lambda_2$ (second eigenvalue of the right Cauchy-Green strain tensor) field (right). Red lines show some contacts (top) and orange bars (vertical at the beginning of the compression) show the rotation of some particles. Pink regions show the voids between grains. e-f: Picture of the soft granular systems in their most strained configuration, $20$\% for the hyperelastic particles (e) and $40$\% for the plastic particles (f).}
\label{Fig_expe}
\end{figure}

One of the most challenging point when dealing with experimental study of highly strained granular matter is to make quantitative measurements at the grain scale for the whole packing. \JB{To overcome this challenge, we} used the novel set-up already presented in \citet{vu2019_em} and shown in fig.~\ref{Fig_expe}a. This consists in a compression machine positioned on a flatbed scanner. Randomly arranged bidisperse soft cylinders lay on a scanner top glass and are compressed step by step in one direction, while imaged from below with the scanner. The so obtained accurate images \JB{($\sim 520$~Mpx)} are analyzed using \JB{the} DIC method \cite{hild2006_str,bornert2009_em,pan2009_mst,vu2019_em} to follow the evolution of the \JB{displacement and strain fields} in each particle.  

As shown in fig.~\ref{Fig_expe}a the compression device is composed of four wall forming \JB{an} inner rectangle of initial dimension $270 \times 202$~mm$^2$. A stepper motor rotate a screw which moves one of the wall along the major direction of the rectangle. Two force sensors are attached to this wall to \JB{hold it and} record the global stress evolution, $\sigma$, when compressing uniaxially the granular system. The global stress is measured continuously with a frequency of $100$~Hz while the system is compressed step by step. At each step the wall moves rightward (along \JB{the} $y$-axis in fig.~\ref{Fig_expe}) of a distance of $0.5$~mm at a speed of $2$~mm/min. Then, the system stays at rest during $1$~min which is enough for the particle to rearrange and the whole \JB{packing} to relax as verified on the stress signal. Once relaxed the packing is imaged from below with a flatbed scanner CanoScan $9000$F Mark II with a resolution of $2400$~dpi ($10.5$~$\mu$m/px). It takes about $10$~min to scan a $8$~bit picture of roughly $19 \times 27$~(kpx)$^2$. Granular systems are compressed up to global strains as high as $40$\%.   

The two most common types of highly deformable material behaviors have been explored in this study; namely hyperelastic incompressible and plastic compressible particles have been studied. The hyperelastic grains are made of silicone \cite{silicone} (see fig.~\ref{Fig_expe}b) whose Young modulus is $E = 0.45$~MPa and Poisson ratio is $\nu = 0.495$ \cite{vu2019_em}. The Poisson ratio very close to $1/2$ provides that the material is almost incompressible. It has been shown that the particle behavior is very close to an ideal hyperelastic one \cite{vu2019_em}. The plastic grains are made of $1$\% agar hydrogel \cite{agar} (see fig.~\ref{Fig_expe}c) which has been shown to behave almost like a perfect elastoplastic material when loaded quasistatically \cite{vu2019_em}. The material parameters are $E = 10$~kPa, \JB{for} the Young modulus, $\nu = 0.15$, \JB{for} the Poisson ratio, $E_p = 1.8$~kPa, \JB{for} the plastic modulus and $\sigma_y = 500$~Pa, \JB{for} the yield stress. The fact that $\nu$ is quite small permits to get highly compressed systems with \JB{a}  packing fraction, $\phi$, larger than $1$.

In both hyperelastic and plastic cases, \JB{bidisperse} grains are casted cylinders of diameters $20$~mm and $30$~mm and of height $15$~mm. \JB{The bidispersity} permits to avoid grain crystallization. For each experiment around $N=100$ particles are used keeping the small over large number of particle ratio constant around $3$. The DIC method to measure displacement fields on the particle bottom faces requires a thin random pattern with a high optical contrast. In the case of hyperelastic particles, this is obtained by mixing the silicone \cite{silicone} with black dye \cite{silicone_black} and coating the mold bottom with a very thin layer of silver glitter \cite{silver_glitter}. In the case of plastic particles, $0.29$\% of black Indian ink and $0.05$\% of thin metallic glitter \cite{silicone_black} in mass are added to the hydrogel before casting. In both cases the glitter characteristic size is $25$~$\mu$m which creates a random pattern with correlation length around $50$~$\mu$m on the bottom face of each particle.

For both sorts of particle three compression runs have been carried out. For each run with hyperelastic particles, a layer of vegetable oil with a low viscous coefficient ($60$~mPa$\times$s) is coated on the glass surface in order to almost \JB{remove} static basal friction and to improve optic transmission. This oil also makes the inter-particle friction vanishing. For plastic particles, a similar effect is obtained by adding DI water in the system. In order to \JB{counter} evaporation of the water contained in the sample all along the experiments, the particles are regularly and gently moistened dropping DI water on the top of them so that they stay saturated in water. Experiments are stopped when particles pop out of the scanner glass due to force chain buckling. This happen after $\sim 20$\% strain for hyperelastic particles (see fig.~\ref{Fig_expe}e) while it reaches twice larger strain values for experiments with plastic particles (see fig.~\ref{Fig_expe}f).

The jamming transition have been widely studied over the past two decades \cite{liu1998_nat,ohern2002_prl,ohern2003pre,katgert2010_epl,zhao2019_arx} and observing it is not the main purpose of this paper. Still to validate the experimental method that we present and even if our system is not large enough for a proper study of this transition, the hyperelastic particle systems have been prepared loose enough to observed this transition point during the compression process. On the contrary to go further in the deeply jammed state, plastic systems have been prepared with a packing fraction very close to the random close packing by densely packing grains by hand at the beginning of the compression. \JB{In both cases, the systems are prepared manually avoiding grain segregation and crystallization as much as possible.}

\subsection{Image processing} \label{imgPro}

For each experiment, sets of $100$ to $200$ black and white pictures, $I_n(x,y)$, are obtained, where $n$ is the compression step going from $0$ to $N$. The thin random patterns induced by the glitter used in the particle making process, permits to perform DIC with an algorithm modified from the one presented in \citet{vu2019_em}. As shown in fig.~\ref{Fig_expe}d this permits to follow the particle displacements, rotations, shapes and strain tensor fields.

The first step of the algorithm consists in detecting the particle positions on the first frame, $I_0(x,y)$. This is done by using the shiny aspect of the particles over the dark background of this picture (see fig.~\ref{Fig_expe}e and f). The image $I_0$ is blurred with a Gaussian noise of standard deviation much larger than the correlation length of the random pattern, namely $300$~$\mu$m. The so obtained image is binarized with a threshold chosen to have $0$ for the background and $1$ for the particle areas. Image convolution is then performed between a pre-set image of a single ideal particle with $3/4$ the diameter of the smallest particles and the binarized image. After performing the convolution, binarizing the resulting image using a threshold of $99$\% of the peak convolution value results in a field of well-isolated regions whose barycenters correspond to the particle centers $\{x^0_i,y^0_i\}_{i \in [0,N]}$. The area of each region indicates the particle's diameter \cite{zadeh2019_arx,all2019_wiki}.

Each particle is then tracked along the compression process by mean of DIC. For a given particle $i$ at the $n^\textrm{th}$ compression step, a sub-image, $I^i_n(x,y)=I_n(x^i_n-100:x^i_n+100,y^i_n-100:y^i_n+100)$, corresponding to a square of $200 \times 200$~px$^2$ centered in the middle of the particle, is extracted from the full picture. The new position of this square in the image of the $(n+1)^\textrm{th}$ step is found using Fourier transform DIC \cite{pan2009_mst}. This consists in convolving $I^i_n(x,y)$ with the full $(n+1)^\textrm{th}$ image $I_{n+1}$ by mean of Fast Fourier Transform (FFT). The maximum of convolution gives the new position of the particle with one pixel accuracy. To get a subpixel accuracy and measure the particle rotation a Nelder-Mead maximization algorithm is applied to the following correlation function:

\begin{equation}
	\mathcal{F}(x_0,y_0,\theta_0)=\sum_{x,y \in \textrm{square}} ( I^i_n(x,y) * \mathcal{T}_{x_0,y_0,\theta_0} \{I_{n+1}\}(x,y) )^2
	\label{eq_DIC}
\end{equation}

\noindent where $\mathcal{T}_{x_0,y_0,\theta_0}\{I\}$ is a function that interpolate a sub-image of $I$ at coordinate $(x_0,y_0)$ with a squared shape of size $200$ tilted of an angle $\theta_0$. The initial guess for the optimization algorithm is the position found with the FFT DIC algorithm and $\theta_0=0$. In very few cases, when important grain rearrangements happen \cite{bares2017_pre} (translation larger than $100$~px or large rotation), the initial guess is given manually by mean of a graphical interface. This permits, for each particle, to obtain a set of smaller images centered around the particle and aligned with the initial orientation of the particle (see orange bars in fig.~\ref{Fig_expe}d). Then, these sets of images where the solid rigid motion has been corrected, can be treated independently as pictures of a complex compression tests on fixed particles.

For each of these sets, the same large displacement DIC algorithm as the one used for a single particle is used \cite{vu2019_em}. A regular grid is defined on the $i^{\textrm{th}}$ particle \JB{at} the first step. This tiling defines correlation cells whose centers are tracked from one image to another to get the particle displacement field $\bm{u}^i_n(x,y)$ (see fig.~\ref{Fig_expe}d). Note that in this case the $(x,y)$ repair is attached to the particle. Just like it is done for the tracking of the particle centers previously described, the center of the correlation cells are tracked by first applying FFT DIC to have a rough estimation of the displacement. This estimation is then improved by maximizing a correlation function similar to the one given in eq.~\ref{eq_DIC}, but neglecting the rotation $\theta_0$. The dimension of the correlation cells are $40$~px ($400$~$\mu$m) which is $8$ times larger than the typical size of the random pattern. This provides a displacement measurement with a strong accuracy: $0.01$~px ($100$~nm). More details about the DIC algorithm are given in \citet{vu2019_em} and \citet{bares2017_hal} \JB{and the Python codes are freely available \cite{DIC_code}}.

\subsection{Local and global measurements} \label{measure}

Many different observables are then, deduced from the displacement field measured at the grain scale. First, the evolution of the deformation gradient tensor, $\bm{F}^i_n(x,y)$, is computed for each particle from $\bm{u}(x,y)$ following the definition \cite{taber2004_bk}: 

\begin{equation}
	\bm{F}= \bm{\nabla}\bm{u} + \bm{I}
	\label{eq_F}
\end{equation}

\noindent with $\bm{I}$ being the second order identity tensor. Similarly the evolution of right Cauchy-Green strain tensor, $\bm{C}^i_n(x,y)$ is obtained from \cite{taber2004_bk}: 

\begin{equation}
	\bm{C} = \bm{F}^T \bm{F}
	\label{eq_C}
\end{equation}

\noindent \JB{Physically, the right Cauchy-Green strain tensor reflects the differences between the metrics of the deformed and undeformed bodies \cite{taber2004_bk}. It is related to the Green-Lagrangian strain tensor $\bm{E}$ using $\bm{E}=1/2(\bm{C}-\bm{I})$. Under the small deformation assumption this last tensor turns out to be the strain tensor $\bm{\varepsilon}$ classically used in elasticity theory. In the large deformation case, $\bm{C}$ is more commonly used than $\bm{E}$ because its principal invariants are directly used to compute energy density functions and constitutive equations \cite{taber2004_bk}. These} different principal invariants are also computed:

\begin{equation}
	\begin{aligned}
		& I_1 = tr(\bm{C}) \\
		& I_2 = 1/2(tr(\bm{C})^2-tr(\bm{C}^2)) \\
		& I_3 = det(\bm{C}) \\
	\end{aligned}
	\label{eq_invar}
\end{equation}

\noindent It is worth noting that in the 2D case $I_2 = I_3$. The two eigenvalues, $\lambda_1$ and $\lambda_2$ (see fig.~\ref{Fig_expe}d), and \JB{corresponding} eigenvectors, $\bm{v}_1$ and $\bm{v}_2$ of the right Cauchy-Green strain tensor are computed as well, along with the von Mises strain field: $\mathcal{C} = \sqrt{\bm{C}:\bm{C}}$. \JB{In this study, since} we compare materials with very different behaviors and elastic moduli, not to make any further assumptions about the exact material behaviors, we do not deal with the stress tensor nor the energy fields. Nevertheless we note that they could be directly deduced from the deformation gradient tensor \cite{taber2004_bk,vu2019_em}.

In the highly strained cases, as shown in fig.~\ref{Fig_expe}e and f, it is not possible to differentiate particle edges near the contact regions. So it is not possible to determine and \JB{track} the particle boundaries directly from the raw images. To follow the evolution of the particle shape, the position of the correlation cells near the particle edges are tracked instead. For a given particle before the first compression step, the centers of the correlation cells belonging to the convex enveloped enclosing all the correlation cells are marked. They form the particle boundary which is followed by tracking the position of these points. A linear interpolation of these points gives the shape function $\rho^i_n(\theta)$ in polar coordinate.

Contacts between particles are detected using these boundaries \JB{and the strain field information. If two particles have their edges closer than a certain threshold ($0.5$~mm) and if, close to the particle boundaries ($1.5$~mm), the von Mises strain $\mathcal{C}$ differs from its average value by more the $1$\% ($|\mathcal{C}-1|>0.01$), then the particles are considered to be in contact. The choice of $1$\% is just above the noise measurement value. This high von Mises strain variation means that locally the deformation is high enough to consider that the contact actually bears a force}. The contact length is directly deduced from this criterion applied locally (see fig.~\ref{Fig_expe}d) and the specific contact length, $l$, is then defined as the ratio between this contact lengths and the particle perimeter.

From the shape function $\rho^i_n(\theta)$ of each particle, we also define the anisotropy, $a_n$, as the ratio between the size of the particle \JB{along} its shortest direction ($\min\limits_{\theta}\{ \rho(\theta)+\rho(\theta+\pi) \}$) over its size \JB{along} its largest direction ($\max\limits_{\theta}\{ \rho(\theta)+\rho(\theta+\pi) \}$). This last major direction is recorded as $\theta_m$. The particle asphericity, $a^p_s$, is also computed from $\rho^i_n(\theta)$ as \cite{boromand2018_prl}:

\begin{equation}
	a^p_s=\frac{p^2}{4 \pi a}
	\label{eq_asphe}
\end{equation}

\noindent where $p$ is the particle perimeter and $a$ is the particle area. So, for each particle $i$ at the $n^\textrm{th}$ compression step, we measure:

\begin{itemize}
	\item the position, $(x^i_n,y^i_n)$; and orientation, $\theta^i_n$,
	\item the displacement field, $\bm{u}^i_n(x,y)$,
	\item the deformation gradient tensor, $\bm{F}^i_n(x,y)$,
	\item the right Cauchy-Green strain tensor, $\bm{C}^i_n(x,y)$; its principal invariants, $(I^i_{1~n}(x,y),I^i_{2~n}(x,y))$; its eigenvalues, $(\lambda^i_{1~n}(x,y),\lambda^i_{2~n}(x,y))$; its eigenvectors, $(\bm{v}^i_{1~n}(x,y),\bm{v}^i_{2~n}(x,y))$; and its von Mises field $\mathcal{C}^i_n(x,y)$,
	\item the shape $\rho^i_n(\theta)$ and the deduced anisotropy, $\{a_n\}^i_n$; asphericity, $\{a^p_s\}^i_n$; and major direction, $\{\theta_m\}^i_n$. 
	\item the contact regions and their lengths.
\end{itemize}

As shown in fig.~\ref{Fig_expe}d, voids surrounding the particles are also \JB{detected}. From the particle positions and shapes, a binarized images are built with $1$ everywhere except in regions covered by particles. Disconnected islands in these pictures form \JB{voids}. To avoid boundary effects only voids that are not in contact with the system edges are kept. Just like for the particles shape, areas, $a_r$, and asphericity, $a^v_s$ (see eq.~\ref{eq_asphe}), are computed. For each void region, the solidity, $s$, is also defined as the ratio between the area of the void and the area of the smallest convex region including this void.

At the global scale, apart from the compression stress, $\sigma$, defined as the measured force over the system width, we also measure the evolution of the packing fraction, $\phi$, the coordination, $Z$, and the fraction of non-rattlers (NR), $f_{\textrm{\scriptsize{NR}}}$. The packing fraction is defined as the ratio between the area of the box prescribing the particles and the sum of the particle areas measured from their initial shape $\rho^i_0(\theta)$. This definition implies that $\phi$ can be greater than $1$ for compressible materials. The coordination, $Z$, is the ratio between the total number of contacts in the system and the number of \JB{NR} particles \cite{goncu2010_crm,bi2011_nat}. NR are particles with at least two force-bearing contacts. The fraction of NR, $f_{\textrm{\scriptsize{NR}}}$, is simply the number of NR particles over the total number of particles.

These quantities have been measured for three experiments with hyperelastic particles and three experiments with plastic particles. \JB{Each time initial configurations were different.} Most of the graphs shown in this paper plot data from two significant experiments, one for each sort of materials. When specified, plots also show results averaged over the \JB{three} experiments.

\section{Results} \label{result}

\subsection{Global observables} \label{sec_global}

\begin{figure}[htb!]
	\begin{center}
		\includegraphics[width=0.49\textwidth]{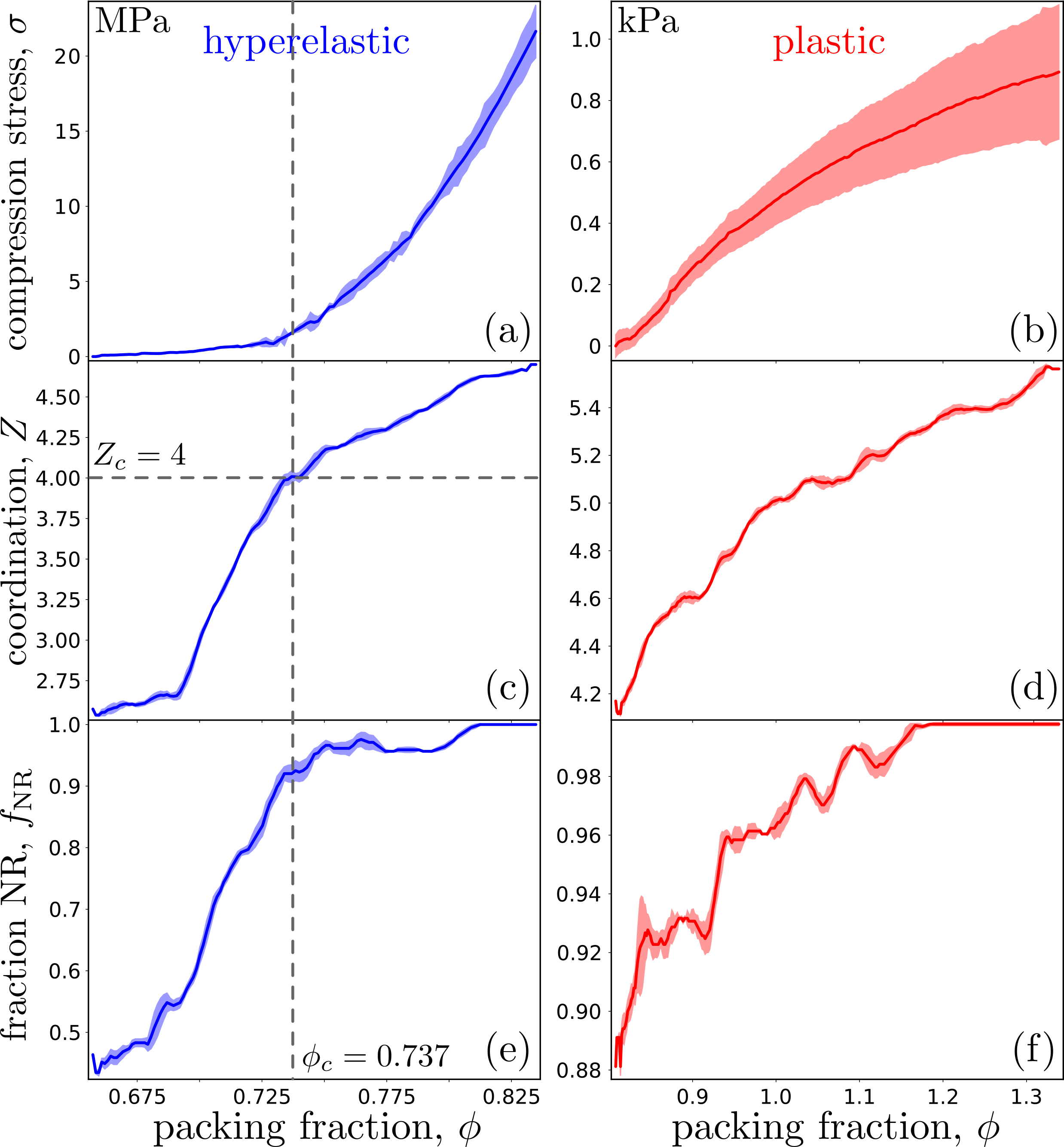}
	\end{center}
\caption{(Color online) Evolution of the compression stress, $\sigma$ (a,b), of the coordination, $Z$ (c,d) and of the fraction of non-rattlers (NR), $f_{\textrm{\scriptsize{NR}}}$ (e,f) as a function of the packing fraction, $\phi$, for hyperelastic (blue curves, left) and plastic (red curves, right) packings. The dashed vertical line shows $\phi_c=0.737$ the critical packing fraction at the jamming point. The horizontal dashed line shows $Z_c=4$, the coordination value for isostatic packing of slippery particles. Results are presented for a single experiment for each material. \JB{They are repeatable from one experiment to another.} The shaded areas correspond with $95$\% error bars. \JB{These error bars are computed by averaging over the force signal for $\sigma$, by averaging over the individual grain coordination for $Z$, and by smoothing the $f_{\textrm{\scriptsize{NR}}}$ over $6$ step for the fraction of non-rattlers.}}
\label{Fig_global}
\end{figure}

For two typical experiments fig.~\ref{Fig_global}a and b show the evolution of the global stress, $\sigma$, as a function of the packing fraction, $\phi$, for hyperelastic and plastic particles respectively. In the first case after a transient regime where the stress stays close to $0$, it rapidly increases with a superlinear regime. On the contrary for plastic particles, $\sigma$ directly enters a sublinear regime. In both cases no saturation regime is observed. The two regimes found for hyperelastic particles correspond with the \textit{unjammed} and \textit{jammed} states \cite{liu1998_nat,ohern2003pre,hecke2009_jpcm}. In the first state, each particle can move without impediment from their neighbors and the bulk modulus is zeros. \JB{This regime is favored by the fact that particles are slippery.} Then, when increasing the packing fraction, \JB{particles cannot significantly rearrange any more and} the system develops a yield stress in a disordered state; it can resist the loading. The local evolution of the packing deep in this last regime is what we mainly investigate in this paper. 

As emphasized in the rest of the fig.~\ref{Fig_global} explained in the next paragraphs, the jamming transition happens at a critical packing fraction $\phi_c \approx 0.737$. \JB{This value significantly varies from one experiment to another ($\pm 0.02$) because it depends on the initial conditions due to the small system size. More importantly, it is lower than the random close packing value} ($\phi=0.842$ \cite{ohern2002_prl}) expected to correspond with the jamming point for frictionless particles isotropically stressed. This low $\phi_c$ value comes from the fact that ($i$) the particles we used are not perfectly frictionless, ($ii$) boundaries are frictional, \JB{($iii$) lubricating liquids induce attractive \JBn{capilary} bridges between particles \cite{halsey1998_prl,mason1999_pre} ($iv$) the compression is uniaxial and ($v$)} the system is quite small so that the jamming packing fraction lays in a broad range \cite{ohern2003pre}. In the plastic case (fig.~\ref{Fig_global}b), the system is prepared with a large initial density and enters the jamming regime at the beginning of the experiment; the unjammed regime is barely not observed.

For the same experiments, in fig.~\ref{Fig_global}c and d, we show the evolution of the coordination number, $Z$, as a function of the packing fraction, $\phi$, for hyperelastic and plastic particles respectively. In the case of hyperelastic particles, after a short regime where $Z$ stays just above $2.5$ contacts per grain it rapidly increases up to $Z=4$ after $\phi \approx 0.692$. Then, it increases \JBn{slowly} in a regime similar to the one observed for plastic particles. This last regime is in agreement with the square root increase already observed in the jammed state for many different experimental systems and numerical simulations \cite{durian1995_prl,ohern2003pre,majmudar2007_prl,katgert2010_epl,winkelmann2017_csa}. However, this finding does not agree the linear regime observed for soap bubbles \cite{winkelmann2017_csa}. \JB{This is most likely due to the fact that in this latter case, particles are not made of a bulk material so their individual mechanical behavior is significantly different of our systems.} The jamming point is crossed for $Z \lesssim 4 = Z_c$ which corresponds with the isostatic point for perfectly slippery particles \cite{hecke2009_jpcm}; above $Z_c$ the system, in the jammed state, is more and more hyperstatic. \JB{One more time, the fact that the jamming point is crossed slightly below $Z=4$ is most likely due to residual friction and the capillary attraction between particles.}

Figure~\ref{Fig_global}e and f show the evolution of the fraction of NR, $f_{\textrm{\scriptsize{NR}}}$, as a function of $\phi$ for the two sorts of particles. In both cases this quantity increases non-monotonously until it reaches a saturation value, $1$, where all the particles are in contact with their neighbors as we can see in fig.~\ref{Fig_expe}e and f. In the case of hyperelastic materials, before the jamming point, $f_{\textrm{\scriptsize{NR}}}$ rapidly increases on the same range as $Z$. Then, it reaches the jamming point at a critical value $f_c~0.9$. The same observation is made for plastic particles. The $f_c$ value is larger than the one measured for frictional particles ($0.83$) jammed by shear \cite{behringer2015_crp}.

Apart from the stress variation in the jammed state, at the global scale, there are no significant differences between the evolution of the systems made of hyperelastic particles and the ones made of plastic ones. In both cases, the fact that curves ($\sigma$, $Z$ and $f_{\textrm{\scriptsize{NR}}}$) are not smooth is explained \JB{by the} sharp grain rearrangements during the compression process reminiscent of what is observed during the shear process \cite{bares2017_pre}. \JB{It is also worth noting that even with the shapes are the same, due to the small system size and the different initial conditions, from one experiment to the other, $\sigma(\phi)$, $Z(\phi)$ and $f_{\textrm{\scriptsize{NR}}}(\phi)$ curves do not collapse.}

\begin{figure*}[htb!]
	\begin{center}
		\includegraphics[width=0.95\textwidth]{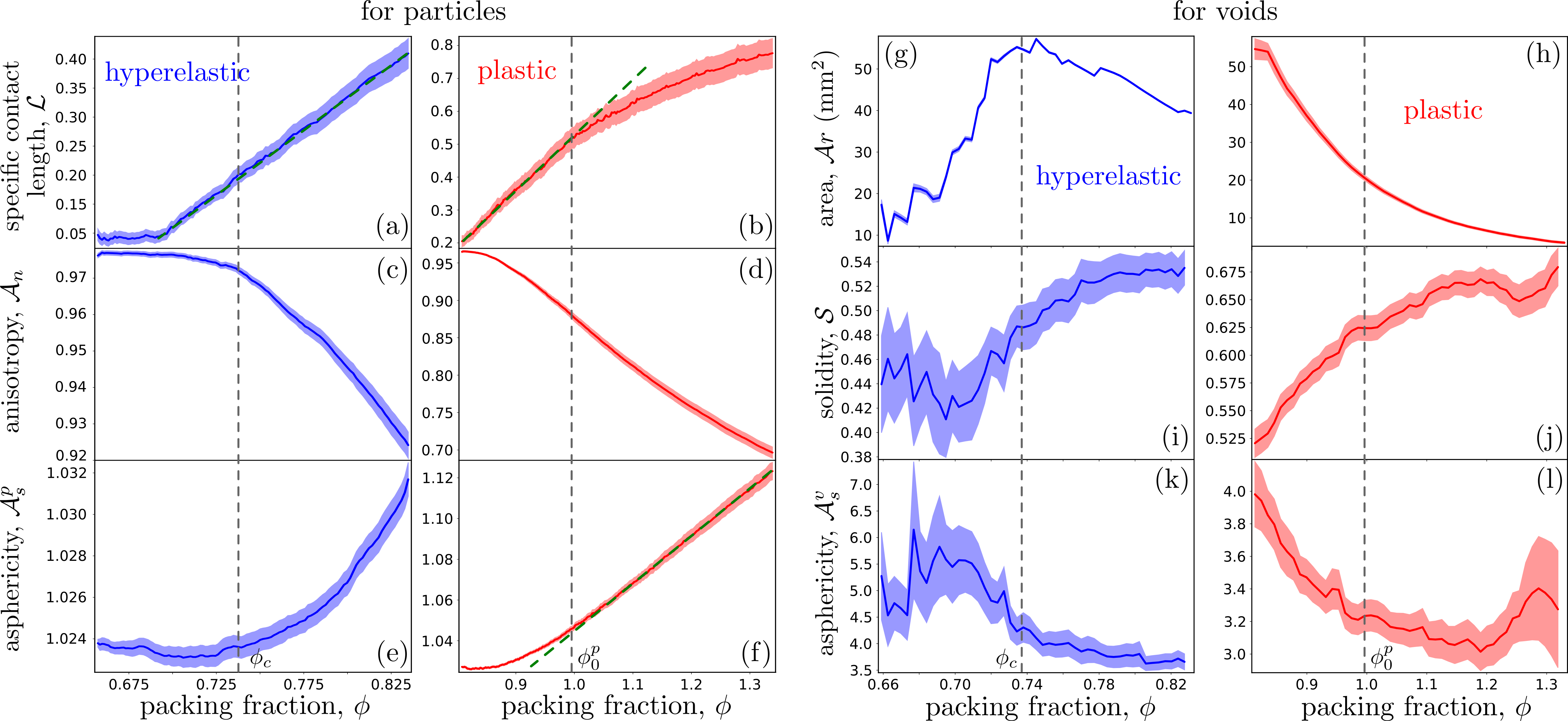}
	\end{center}
\caption{(Color online) Evolution of the average inter-particle contact specific length, $\mathcal{L}$ (a, b), of the average particle shape anisotropy, $\mathcal{A}_n$ (c, d) and asphericity, $\mathcal{A}^{p}_s$ (e, f) as a function of the packing fraction, $\phi$, for hyperelastic (blue curves, left) and plastic (red curves, right) packings. In a, the green dashed line \JB{shows} a linear fit with slope $2.63$, in b it shows a slope of $1.23$ and in f it shows a slope of $0.23$. Evolution of the total void area, $\mathcal{A}r$ (g, h), of the average void shape solidity, $\mathcal{S}$ (i, j) and asphericity, $\mathcal{A}^{v}_s$ (k, l) as a function of the packing fraction $\phi$ for hyperelastic (blue curves, left) and plastic (red curves, right) packings. Quantities are averaged over all the voids detected in the center of the packing. Results are presented for a single experiment. The dashed vertical \JB{lines} in hyperelastic particle graphs shows $\phi_c=0.737$ the critical packing fraction at the jamming point. The dashed vertical \JB{lines} in plastic particle graphs shows $\phi^p_0 \approx 0.99$ a crossover packing fraction. The shaded areas correspond with $95$\% error bars \JB{averaging over particles or voids}.}
\label{Fig_global_edge}
\end{figure*}

For the same typical experiments as the ones used for fig.~\ref{Fig_global}, in fig.~\ref{Fig_global_edge}a and b, we show the evolution of the average specific contact length, $\mathcal{L}=\left\langle l \right\rangle $, during the compression process for hyperelastic and plastic particles respectively. In the hyperelastic case after a short transient regime where $\mathcal{L}$ stays below the noise measurement level, it then increases linearly with slope $\sim 2.63$ just before the jamming transition \JB{point}. The beginning of the linear increase ($\phi \approx 0.692$) corresponds with the growth of the coordination in fig.~\ref{Fig_global}c. It corresponds with the beginning of the force chain building up due to finite size effects \JB{(arches in the system)} and residual basal friction. In the plastic particle case, this increase is also linear with slope $\sim 1.23$ until a crossover around $\phi = \phi^p_0 \approx 0.99$; then it becomes sublinear. \JB{Contrary to $\phi_c$, the value of this crossover $\phi^p_0$ is stable from one experiment to another, it does not significantly depend on the initial conditions.}

The particle average anisotropy, $\mathcal{A}_n = \langle a_n \rangle$ and the asphericity $\mathcal{A}^p_s=\langle a^p_s \rangle$ are plotted as a function of the packing fraction for hyperelastic and plastic particles in fig.~\ref{Fig_global_edge}c to f respectively. For hyperelastic particles, $\mathcal{A}_n$ and $\mathcal{A}^p_s$ stay close to $1$ before jamming ($\phi < \phi_c$) and then rapidly decrease and increase respectively after the jamming point. This evidences an important and continuous deformation of the particles from their initial circular shape. Both values are not exactly $1$ in the undeformed configuration since they are computed using a polygonal interpolation ($\rho(\theta)$) of the circular particles. For plastic particles, $\mathcal{A}_n$ first slowly decreases and then keeps on decreasing sublinearly. For $\mathcal{A}^p_s$, the curve progressively increases and enters a linear regime with slope $0.23$ from the crossover point $\phi = \phi^p_0$. We note that around the jamming point $\mathcal{A}^p_s(\phi)$ follows the same tendency as the one evidenced by \citet{boromand2018_prl} for numerical simulations of squeezed polygons and that the maximum asphericity ($1.22$) is also close to the maximum they found ($1.25$).	

Figures~\ref{Fig_global_edge}g and h show the evolution of the total void area $\mathcal{A}_r = \sum a_r$, as a function of $\phi$ for hyperelastic and plastic particles respectively. In the first case the total void area erratically increases \JB{up} to the jamming point were almost all the interstitial voids are formed thanks to the particle rearrangements and \JB{induced} contact nucleations. Then $\mathcal{A}_r$ decreases due to the particle deformation and the flow of the particle matter toward the voids. In the plastic case the system is very close to jamming at the beginning and most the voids are already formed. So it directly and sharply \JB{decreases} in a \JB{sublinear} manner. 

Figures~\ref{Fig_global_edge}g to l show the evolution of the average void solidity $\mathcal{S} = \langle s \rangle$ and asphericity $\mathcal{A}^v_s = \langle a^v_s \rangle$, as a function of the packing fraction for hyperelastic and plastic particles respectively. In the case of hyperelastic particles, both $\mathcal{S}$ and $\mathcal{A}^v_s$ are constant within their error bars until $\phi$ reaches the point where the coordination and the fraction of NR begin to \JB{increase} ($\phi \approx 0.692$ in fig.~\ref{Fig_global}c and e). Then, $\mathcal{S}$ increases and $\mathcal{A}^v_s$ decreases until they both reach a plateau value. In the case of plastic particles $\mathcal{S}$ and $\mathcal{A}^v_s$ directly increases and decreases respectively and then both plateau around a constant value. The plateau values correspond with the fact that the void are formed and their shapes evolve only homothetically while their areas decrease.

\subsection{Local observables statistics} \label{sec_local}

\subsubsection{Particle geometry}

\begin{figure}[htb!]
	\begin{center}
		\includegraphics[width=0.45\textwidth]{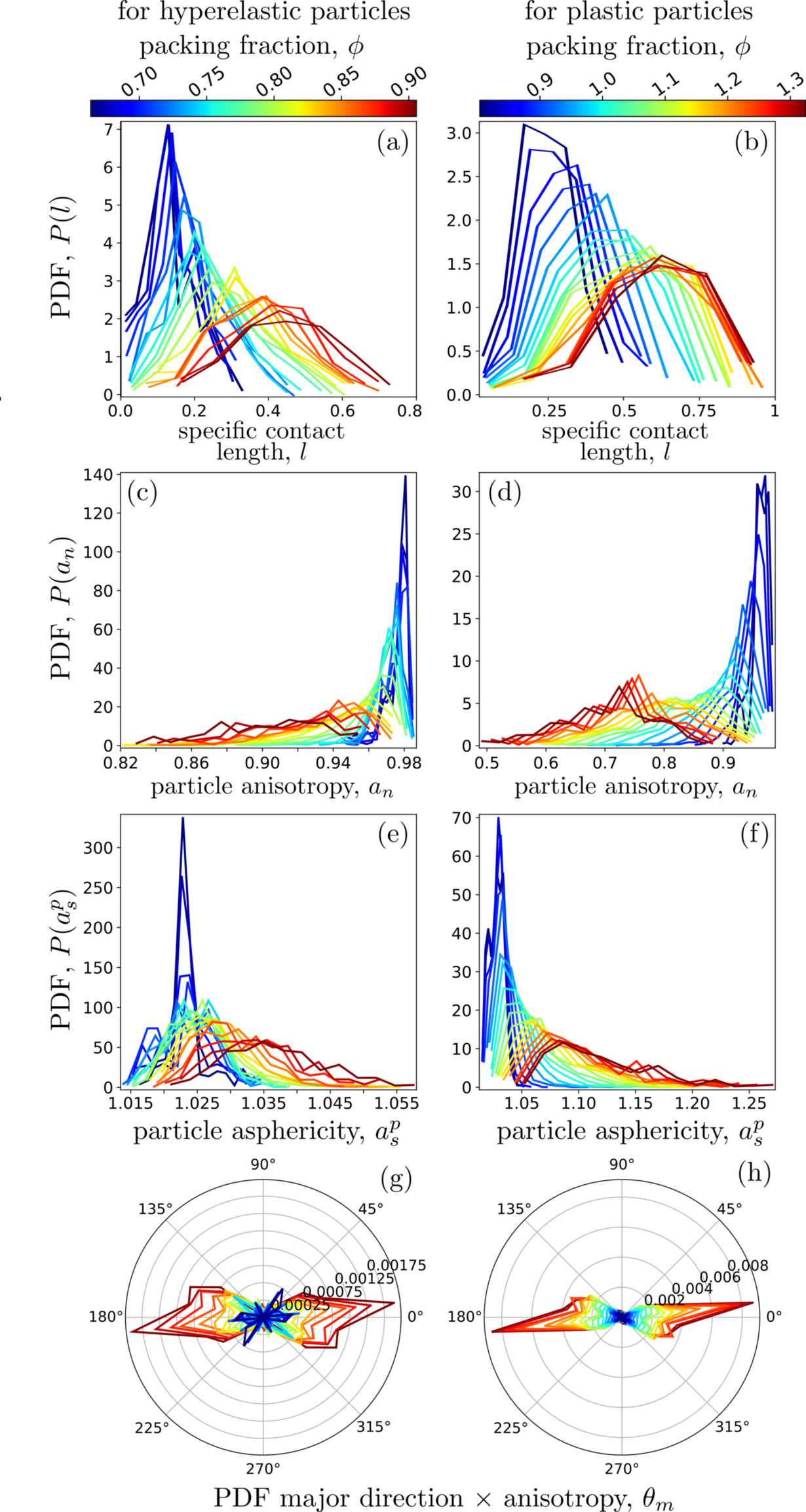}
	\end{center}
\caption{(Color online) a, b: Evolution of the probability density function (PDF) of the specific contact length per particle, $P(l)$, as a function of the packing fraction, $\phi$, for hyperelastic and plastic grains respectively. c, d: Evolution of the probability density function of the particle anisotropy, $P(a_n)$, as a function of the packing fraction, $\phi$, for hyperelastic and plastic grains respectively. e, f: Evolution of the PDF of the particle asphericity, $P(a_s^p)$, as a function of the packing fraction, $\phi$, for hyperelastic and plastic grains respectively. g, h: Evolution of the PDF of the particle major orientation, $\theta_m$, times the anisotropy as a function of the packing fraction, $\phi$, for hyperelastic and plastic grains respectively. Packing fraction color code is given at the top of the left column for hyperelastic grains and of the right column for plastic ones. Results are average over three runs for each material.}
\label{Fig_PdfGrain}
\end{figure}

Figures~\ref{Fig_PdfGrain}a and b show how the probability density function (PDF) of the specific contact lengths, $l$, evolves with $\phi$ for experiments with hyperelastic and plastic particles respectively. These lengths $l$ are computed for each particle and PDFs are averaged over three experiments for each material. For both materials, the PDFs are Gaussian-like with increasing mean and standard deviation values. \JB{This means that more and more contacts are created and also that their individual length increases with a broader and broader diversity in size.} In the plastic case for $\phi>1.15$, the curves collapse which means that the contact network of the packing is frozen. \JB{In this case the individual contact lengths barely increase, the evolution of the system is mainly due to the compressibility of the particle material. This is in agreement with the fact for $\phi>1.15$ void shapes do not evolve as seen in fig.~\ref{Fig_global_edge}h, j and l.}

In fig.~\ref{Fig_PdfGrain}c and d we show the evolution of the particle anisotropy PDF, $P(a_n)$, as a function of the packing fraction $\phi$ for experiments with hyperelastic and plastic particles respectively. These anisotropies $a_n$ are computed for each particle and PDFs are averaged over three experiments for each material. At first the particle anisotropies are all very close to $1$ so $P(a_n)$ is a narrow Gaussian centered around this value. This corresponds this the fact that particles are \JBn{undeformed.} When compressing, for both materials, the average value decreases while the standard deviation rapidly increases. This means that particles are deformed with a broad range of shape diversity, the ones involved in strong force chains being more deformed than the others. In the case of plastic particles, for highly deformed states -- typically $\phi>1.15$ -- a broad Gaussian-like peak re-appears which means that most of the particles are deformed with the same manner; they form polygons with sharp edges as shown in fig.~\ref{Fig_expe}f.

Figures~\ref{Fig_PdfGrain}e and f show the evolution of the grain asphericity PDF, $P(a^p_s)$, for experiments with hyperelastic and plastic particles respectively. These quantities, $a^p_s$, are computed for each particle and PDFs are averaged over three experiments for each material. As for the anisotropy, at the beginning, the particle asphericities are concentrated around $1$\JB{, when particles are not deformed}. Then when the compression level increases, average value increases as well as standard deviation. However, unlike for $l$ and $a_n$ no saturation regime is observed for plastic particles, $P(a^p_s)$ keeps on evolving even at high packing fraction. 

In fig.~\ref{Fig_PdfGrain}g and h we show the evolution of the particle major direction, $\theta_m$, distribution, as a function of the packing fraction $\phi$ for experiments with hyperelastic and plastic particles respectively. By sake of clarity, the major direction of each particle is multiplied by its anisotropy before computing the PDFs. Also, PDFs are averaged over three experiments for each material. For both materials, at the beginning of the compression no preferred direction appear for the particle geometry, the behavior is reminiscent of an isotropic compression. This is due to the fact that particle can rearrange to homogenize the strain \JB{in the system}. However, after few percent of deformation, a preferred direction emerges and amplifies; particles flatten long the normal direction to the compression axis. \JB{As the compression strain level is higher for plastic particles, the major direction PDF has a larger amplitude.}

\subsubsection{Void geometry}

\begin{figure}[htb!]
	\begin{center}
		\includegraphics[width=0.45\textwidth]{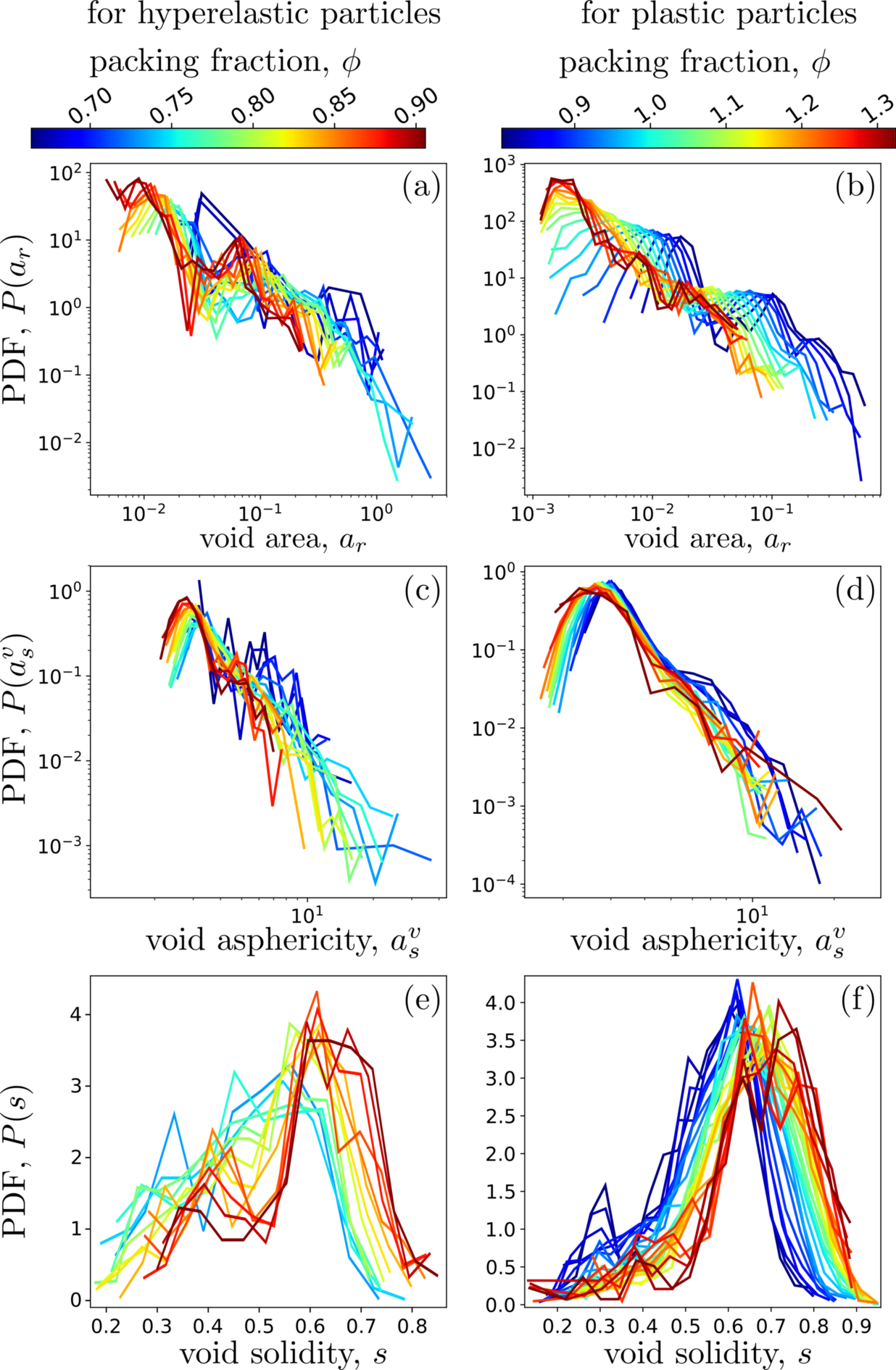}
	\end{center}
\caption{(Color online) a, b: Evolution of the probability density function (PDF) of the scaled void area, $P(a_r)$, as a function of the packing fraction, $\phi$, for hyperelastic and plastic grains respectively. The area, $a_r$, is scaled with the area of the largest particles. c, d: Evolution of the PDF of the void asphericity, $P(a_s^v)$, as a function of the packing fraction, $\phi$, for hyperelastic and plastic grains respectively. e, f: Evolution of the PDF of the void solidity $P(s)$ as a function of the packing fraction, $\phi$, for hyperelastic and plastic grains respectively. Packing fraction color code is given at the top of the left column for hyperelastic grains and of the right column for plastic ones. Results are average over three runs for each material. The four first graphs are in log-log scale.}
\label{Fig_PdfVoid}
\end{figure}

In fig.~\ref{Fig_PdfVoid} we show the evolution of the PDFs of the different void geometry observables during the compression process. Figures~\ref{Fig_PdfVoid}a and b plot the PDFs of the scaled void areas, $P(a_r)$, for hyperelastic and plastic particles respectively. These PDFs span several orders of magnitude and are aligned along power-laws that shift toward smaller relative areas when the packing fraction increase. Figures~\ref{Fig_PdfVoid}c and d present the PDFs of the void asphericities, $P(a^v_s)$, when $\phi$ increases for the two same experiments. In both hyperelastic and plastic cases PDFs collapse on a master curve. Most of the voids have an asphericity value around $3$ and a power-law tail describe the values of the remaining ones. Figures~\ref{Fig_PdfVoid}e and f show the PDF of the void solidity, $P(s)$, for increasing packing fraction for hyperelastic and plastic experiments respectively. In both cases most of the voids have a solidity value around $0.6$ and this value slightly increases with the packing fraction. Whatever is the observable there is no significant differences between the evolution of the void geometries in the hyperelastic and the plastic cases.

\subsection{Strain field evolution} \label{sec_strain}

Beyond the geometrical aspects, the DIC at the grain scale gives information about the local strain \JB{in the system}. Below the jamming point, this strain stays negligible; even if it does not exactly vanish due \JB{mainly} to boundary effects -- force chains along walls for example. On the contrary, above the jamming point in the highly deformed regime, the strain field can reach high levels and becomes a relevant quantity to study the granular packing evolutions. 

\begin{figure}[htb!]
	\begin{center}
		\includegraphics[width=0.4\textwidth]{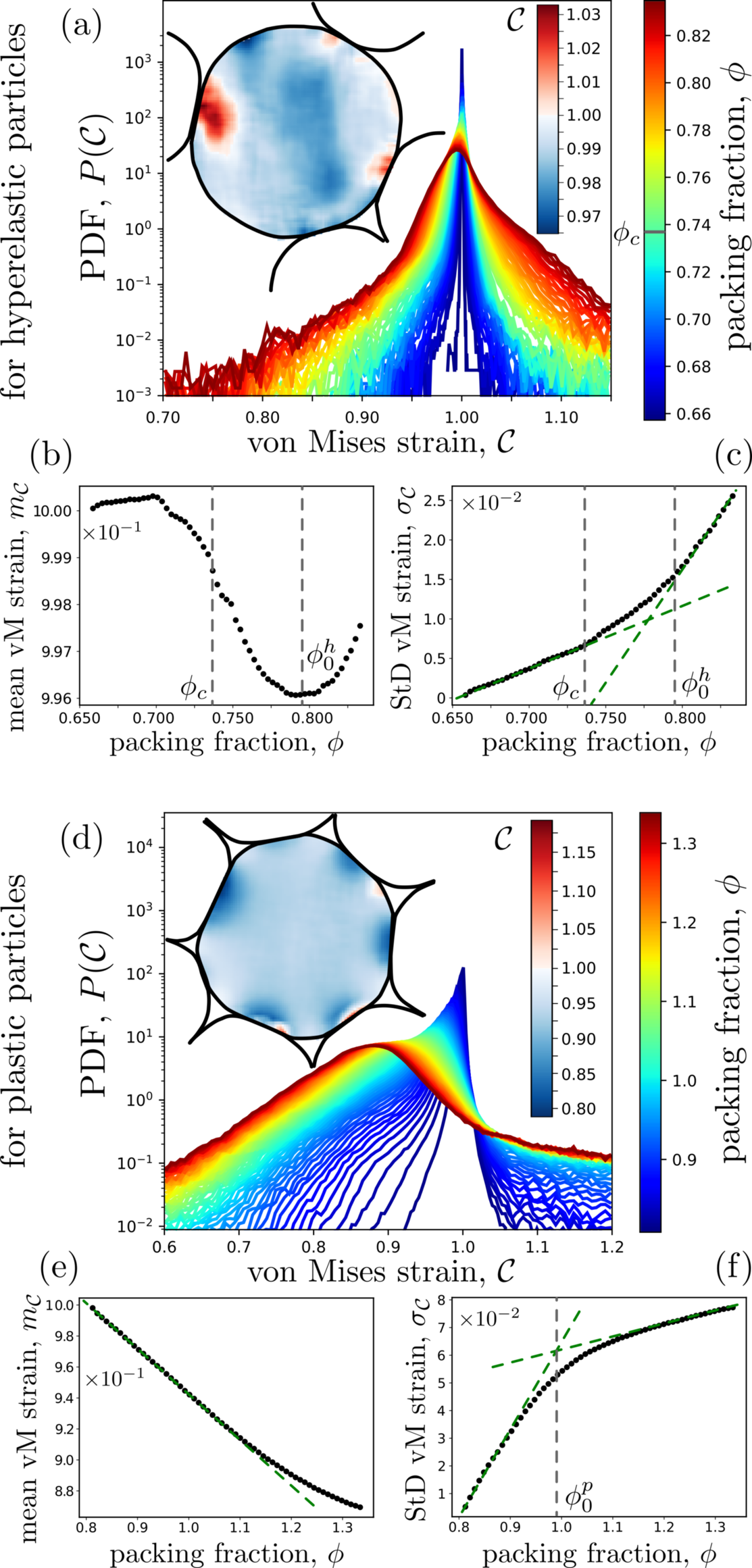}
	\end{center}
\caption{(Color online) Evolution of the probability density function (PDF) of the von Misses strain inside the grains, $P(\mathcal{C})$, as a function of the packing fraction, $\phi$, for hyperelastic (a) and plastic grains (d). \JB{Inset (a) and (d): von Mises strain field for large hyperelastic and plastic particles respectively. The measure is made for a packing fraction close to $\phi_0$. Color bars are from blue to red.} Evolution of the von Misses strain mean value, $m_{\mathcal{C}}$ (b, e) and standard deviation, $\sigma_{\mathcal{C}}$ (c,f) as a function of the packing fraction. Results for hyperelastic grains are given in b and c. Results for plastic grains are given in e and f. Results are presented for a single experiment. In c, the green dashed lines show slope of $0.08$ and $0.28$. In e, it shows a slope of $-0.30$ and in f slopes are $0.31$ and $0.05$. In b and c vertical dashed lines show the packing fraction $\phi_c \approx 0.737$ and $\phi^h_0 \approx 0.795$. In f it shows $\phi^p_0 \approx 0.99$.}
\label{Fig_PdfStrainVM}
\end{figure}

Figure~\ref{Fig_PdfStrainVM}a shows the evolution of the \JB{the von Mises strain} $\mathcal{C}(x,y)$ PDF measured for all the particles of a typical experiment carried out with hyperelastic material. At the beginning of the compression, $P(\mathcal{C})$ \JB{is} a sharp Gaussian centered around $1$ with a \JB{very narrow} standard deviation \JB{induced} by the noise measurement level. \JB{This almost constant value is due to the fact that the whole system is unloaded so no deformation is measured and $\mathcal{C}=1$.} For higher packing fraction the distribution \JB{shifts to the left and} gets broader and broader. \JB{This left shifting comes from the fact that the system is continuously compressed so the von Mises strain gets smaller than $1$. However, as shown in the inset of fig.~\ref{Fig_PdfStrainVM}a, for areas near the strongly loaded edges, instead of being smaller than $1$, $\mathcal{C}$ is slightly greater. This comes \JBn{from} the material incompressibility: instead of being only squeezed, in Eulerian coordinates, matter is pushed apart normally to the compression direction making the right Cauchy-Green strain larger than $1$. This is in agreement with what is shown in fig.~\ref{Fig_PdfStrainVM}b. The mean value of the $P(\mathcal{C})$ distribution, $m_{\mathcal{C}}$,} stays first close to $1$ and then begins to decrease rapidly around the jamming point -- $\phi=\phi_c$. \JB{This corresponds with the fact that just before jamming the system forms weak force chains percolating only in the compression direction so $m_{\mathcal{C}}$ slowly begins to decrease before jamming. At the jamming point, force chains percolate in every direction so compression strain increases in more and more particles and $m_{\mathcal{C}}$ decreases on average.} It reaches a minimum for $\phi^h_0 \approx 0.795$ and then increases back. This corresponds to the point where material dilation due to matter incompressibility begins to grow faster than material compression (see field in the inset of \JBn{fig.~\ref{Fig_PdfStrainVM}a).}

\JB{This is also in agreement with fig.~\ref{Fig_PdfStrainVM}c showing} the evolution of the von Mises strain standard deviation, $\sigma_{\mathcal{C}}$. This quantity continuously increases. Below the jamming point and above $\phi^h_0$, $\sigma_{\mathcal{C}}$ grows linearly with $\phi$ with slopes $0.08$ and $0.28$ respectively. In between a smooth crossover is observed. \JB{It corresponds with the fact that matter dilation increases faster making the standard deviation of $\mathcal{C}$ getting larger and larger. This also explains why in log-scale the PDF tails of $P(\mathcal{C})$ appears asymmetrical: There is no reason for the dilation and compression areas to have the same statistical distribution. We note here that this asymmetricality is very gentle and only seen in log-scale. So even if not shown here for the sake of conciseness, the skewness of $\mathcal{C}$ does not change significantly around $\phi_c$ and $\phi^h_0$.}

In fig.~\ref{Fig_PdfStrainVM}d, we plot $P(\mathcal{C})$ \JB{when $\phi$ increases} for a typical experiment with plastic particles. As observed for hyperelastic particles, it first follows a sharp Gaussian centered around $1$. However, it differs for highly stained states since the peak position rapidly \JB{decreases to} the left hand side. \JB{Contrary to what is observed for hyperelastic material, this peak never turns back to higher $\mathcal{C}$ values because of the material compressibility. As shown in the inset of fig.~\ref{Fig_PdfStrainVM}d, except in very narrow areas close to the most loaded contacts where the incompressibility assumption is less pertinent, the strain is always smaller than $1$: Material compressibility does not induce matter rearrangements normally to the compression direction. The monotonous shift of the peak is also evidenced by fig.~\ref{Fig_PdfStrainVM}e, that shows the mean von Mises strain, $m_{\mathcal{C}}$, decreasing} linearly with $\phi$ (slope $-0.30$) before saturating for packing fractions larger than $1.2$. On the contrary, the standard deviation, $\sigma_{\mathcal{C}}$, increases first linearly with slope $0.31$ and after a crossover centered around $\phi^p_0 \approx 0.99$ increases with a less stiff slope ($0.05$). This crossover deep in the jammed regime around $\phi^p_0$ evidences a microscopic change in the granular matter behavior. \JB{One more time it could be explained by the competition between compression and dilation, even if this latter is less strong than the one observed from hyperelastic material. In the plastic case, PDF tails are also asymmetrical, with almost not values larger than $1$ and an exponential left hand side tail.}

\begin{figure}[htb!]
	\begin{center}
		\includegraphics[width=0.4\textwidth]{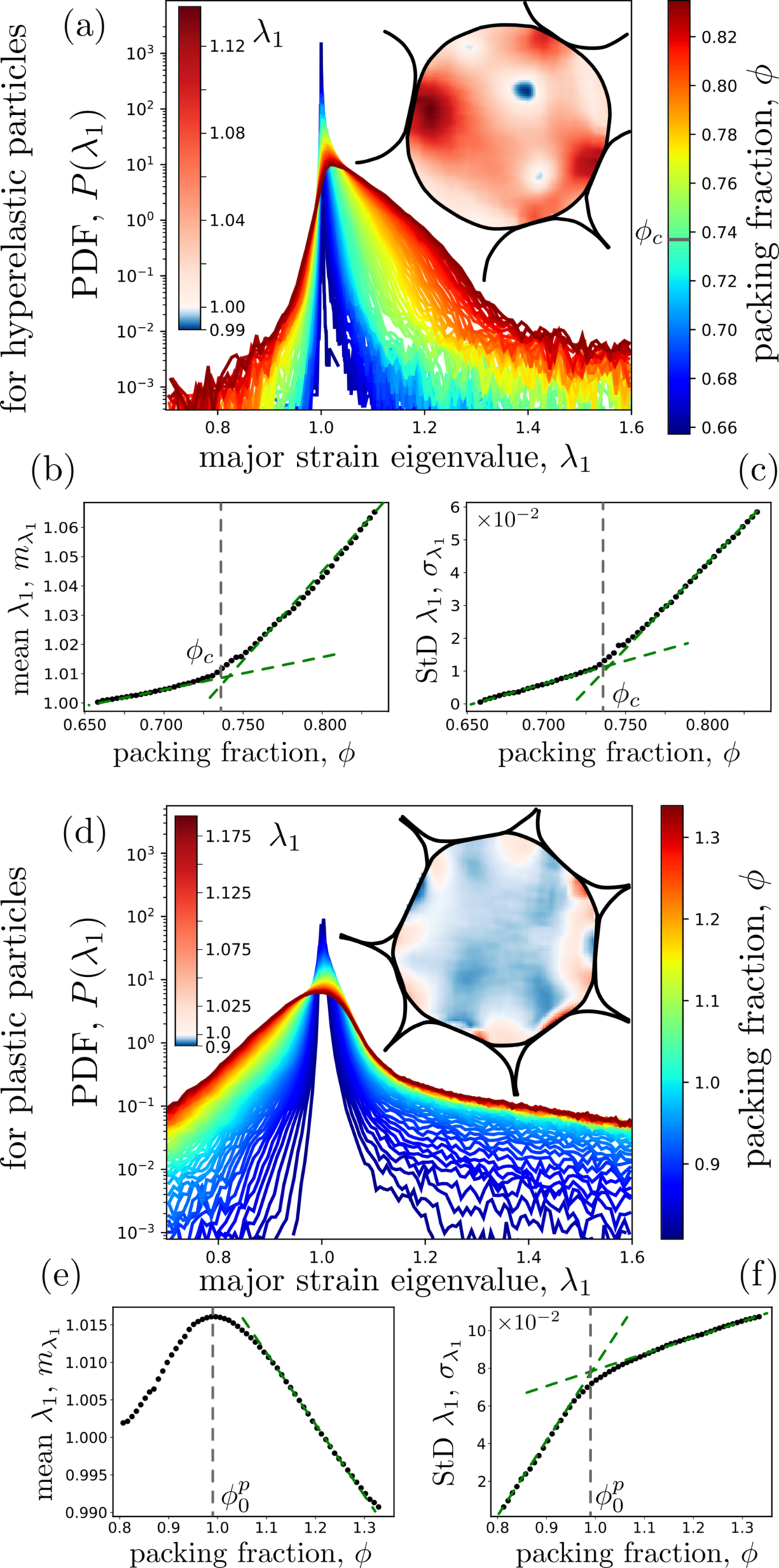}
	\end{center}
\caption{(Color online) Evolution of the probability density function (PDF) of the major strain eigenvalue, $P(\lambda_1)$, as a function of the packing fraction, $\phi$, for hyperelastic (a) and plastic grains (d). \JB{Inset (a) and (d): major strain eigenvalue field for large hyperelastic and plastic particles respectively. The measure is made for a packing fraction close to $\phi_0$. Color bars are from blue to red.} Evolution of the average value of $\lambda_1$, $m_{\lambda_1}$ (b, \JB{e}) and of its standard deviation, $\sigma_{\lambda_1}$ (c,f) as a function of the packing fraction. Results for hyperelastic grains are given in b and c. Results for plastic grains are given in e and f. Linear fits are shown in the different graphs with green dashed lines. Slopes are $0.12$ and $0.54$ in b, $0.14$ and $0.54$\JB{verif} in c, $0.094$ in e, and $0.39$ and $0.087$ in f. In b and c vertical dashed lines show the packing fraction $\phi_c \approx 0.737$ and e and f it shows $\phi^p_0 \approx 0.99$. Results are presented for a single experiment for each material.}
\label{Fig_PdfStrainl1}
\end{figure}

In fig.~\ref{Fig_PdfStrainl1}a, we show the PDFs of the first eigenvalue of the right Cauchy-Green strain tensor,$\lambda_1$, for increasing packing fractions when compressing hyperelastic particles. As observed for the von Mises strain in fig.~\ref{Fig_PdfStrainVM}a, for \JB{weakly} strained systems, $P(\lambda_1)$ is a narrow Gaussian distribution \JB{centered around $1$ since there is almost no deformation. As shown in the inset of fig.~\ref{Fig_PdfStrainl1}a, $\lambda_1$ increases where the matter is loaded and stays close to $1$ where the system remains unstrained. This implies that}, when $\phi$ increases, the mean value of $\lambda_1$, $m_{\lambda_1}$, and its standard derivation, $\sigma_{\lambda_1}$ grow (see fig.~\ref{Fig_PdfStrainl1}b and c). \JB{Both increase} linearly in the unjammed and jammed regimes \JB{with a crossover in between, at $\phi_c$.} We note that we do not see any inflexion of the curves near $\phi=\phi^h_0$. \JB{We believe this is due to the fact that $\lambda_1$ is not a sensitive marker of the competition between compression and dilation of the hyperelastic matter.}

Figure~\ref{Fig_AvgShape}d show the evolution of $P(\lambda_1)$ during the compression of plastic particles. \JB{Similarly to what is observed for hyperelastic particles, t}he $P(\lambda_1)$ goes from a sharp Gaussian \JB{centered around $1$} at low compression level to a broader distribution with an exponential tail at low $\lambda_1$ and a rapidly then slowly decreasing value for high $\lambda_1$. \JB{The inset of \ref{Fig_AvgShape}d shows the $\lambda_1$ field of a particle for a packing fraction near $\phi^p_0$. Its value seems to be higher in areas where matter is sheared and close to $1$ elsewhere.} As shown in fig.~\ref{Fig_AvgShape}e, the average value of $\lambda_1$ first increases with $\phi$ until it reaches a maximum value for $\phi = \phi^p_0$. It then decreases linearly with slope $0.094$. \JB{We believe this maximum value corresponds to a point where on average, the particles are so deformed that isotropic compression dominate over shear.} The standard deviation of $\lambda_1$ whose evolution with $\phi$ is shown in fig.~\ref{Fig_AvgShape}f follows the same tendency as $\sigma_{\mathcal{C}}$ presented in fig.~\ref{Fig_PdfStrainVM}f; it follows two different linear regimes with slopes $0.39$ and $0.087$ separated by a smooth crossover at $\phi = \phi^p_0$.  

\JB{For both hyperelastic and plastic particles, we observe asymmetric PDF tails, that are not related to the skewness, when the systems are squeezed. This is due to the intrinsic asymmetrical nature of the eigenvalues. We also note that the same analysis} has been done for the second and complementary eigenvalue of $\bm{C}$, $\lambda_2$. We have found curves that are symmetric with respect to the axis \JB{$\lambda=1$} and made similar conclusions.

\begin{figure}[htb!]
	\begin{center}
		\includegraphics[width=0.4\textwidth]{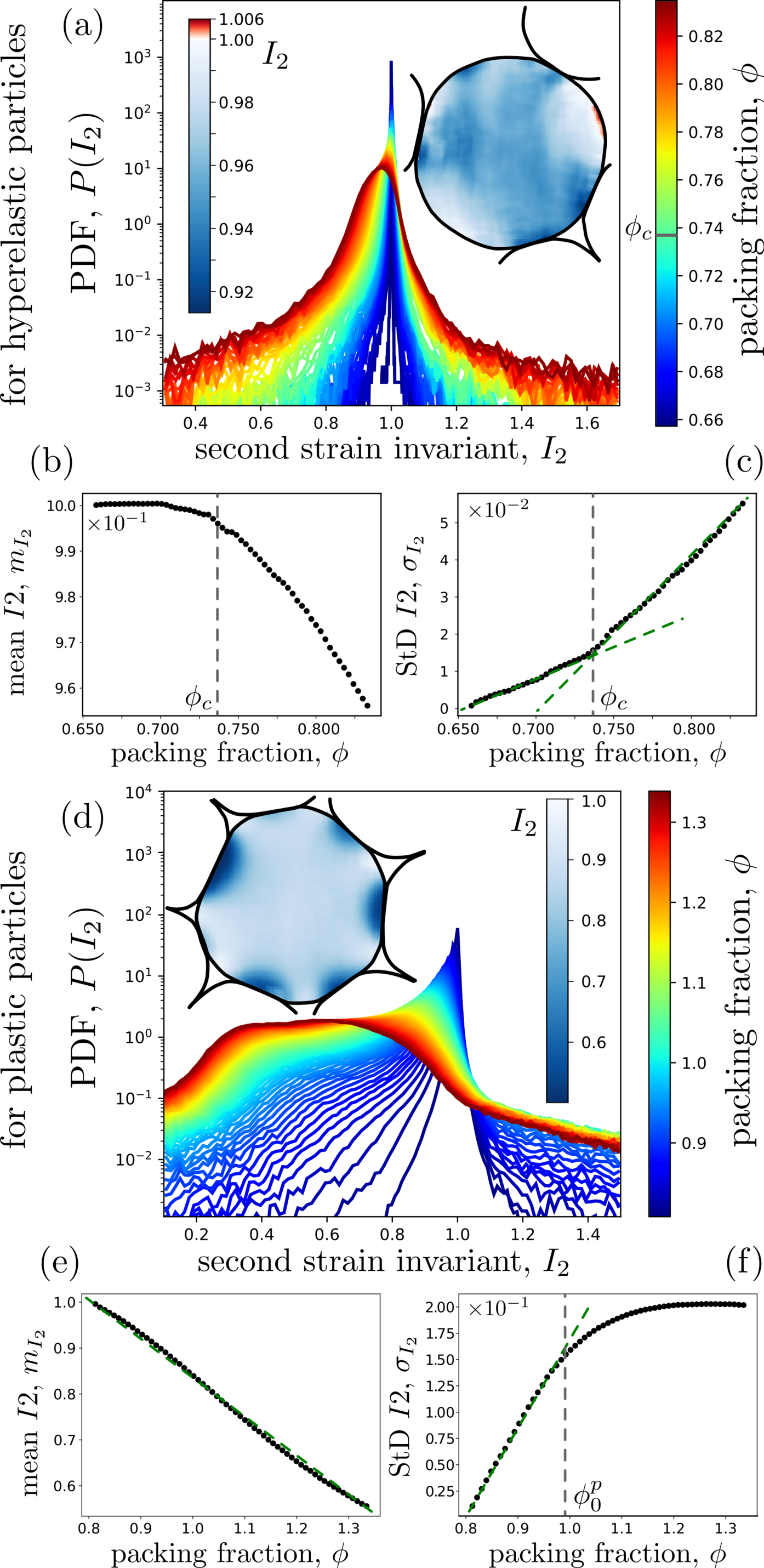}
	\end{center}
\caption{(Color online) Evolution of the probability density function (PDF) of the second strain principal invariant, $P(I_2)$, as a function of the packing fraction, $\phi$, for hyperelastic (a) and plastic grains (d). \JB{Inset (a) and (d): second strain principal invariant field for large hyperelastic and plastic particles respectively. The measure is made for a packing fraction close to $\phi_0$. Color bars are from blue to red.} Evolution of the average value of $I_2$, $m_{I_2}$ (b, \JB{e}) and of its standard deviation, $\sigma_{I_2}$ (c,f) as a function of the packing fraction. Results for hyperelastic grains are given in b and c. Results for plastic grains are given in e and f. Linear fits are shown in the different graphs with green dashed lines. For c slopes are $0.18$ and $0.42$, for e it is $0.84$ and for f it is $0.83$. In b and c vertical dashed lines show the packing fraction $\phi_c \approx 0.737$ and f it shows $\phi^p_0 \approx 0.99$. Results are presented for a single experiment for each material.}
\label{Fig_PdfStrainI2}
\end{figure}

Figure~\ref{Fig_PdfStrainI2}a shows the evolution of the PDF of the second principal invariant of $\mathcal{C}$ measured for all the particles of a typical experiment carried out with hyperelastic material. For low packing fraction, $P(I_2)$ is a narrow Gaussian centered around $1$. When $\phi$ increases this mean value decreases but the PDF stays symmetric contrary to what has been observed for $P(\mathcal{C})$ and $P(\lambda_1)$. \JB{In 2D, $I_2 = I_3 = \textrm{det}(\bm{F})^2$ which is the square of the dilation ratio \cite{taber2004_bk} in Lagrangian coordinates. So, as shown in the in set of fig.~\ref{Fig_PdfStrainI2}a, $I_2<1$ where the particle is squeezed, $I_2 = 1$ where there is no deformation and $I_2>1$ where material expands. This is in agreement with what is shown in fig.~\ref{Fig_PdfStrainI2}b and c.} The evolution of mean and standard deviation of $I_2$, $m_{I_2}$ and $\sigma_{I_2}$ respectively \JB{display two different regimes}. In the unjammed state, $m_{I_2}$ stays constant and $\sigma_{I_2}$ linearly increases with a slop of $0.18$. After a crossover around the jamming point ($\phi = \phi_c$), $m_{I_2}$ rapidly decreases and $\sigma_{I_2}$ increases linearly with a steeper slope ($0.42$). \JB{The decrease of $m_{I_2}$ after $\phi_c$ corresponds with the global squeezing of the system while the increase of $\sigma_{I_2}$ corresponds with the fact that particle material is more and more split between compression and expansion areas.}

In fig.~\ref{Fig_PdfStrainI2}d, we plot $P(I_2)$ for increasing $\phi$ for a typical experiment with plastic particles. Like for hyperelastic particles, it first follows a sharp Gaussian regime centered around $1$. When the packing fraction increases, the $P(I_2)$ peak shift to the left and becomes wider and wider until $P(I_2)$ displays a broad plateau between $I_2=0.3$ and $I_2=0.7$. \JB{As shown in the inset of fig.~\ref{Fig_PdfStrainI2}d, due to the almost perfect particle compressibility, $I_2$ is lower than $1$ everywhere and smaller close to the contact points where the squeezing is maximum. Figure~\ref{Fig_PdfStrainI2}e shows that,} surprisingly the mean second invariant value evolves linearly with $\phi$ decreasing with slope $0.84$. No inflection is observed around $\phi = \phi^p_0$. However, as shown in fig.~\ref{Fig_PdfStrainI2}f, below $\phi^p_0$ the standard deviation of $I_2$ increases linearly with slope $0.83$ while it plateaus above. One more time, this evidences a microscopic change in the granular matter behavior. \JB{We believe it is explained by the fact that above $\phi^p_0$ the system is so deformed that there are almost no void anymore and the matter deforms homogeneously in the compression direction just like a bulk material would do. We also note that the plateau observed in fig.~\ref{Fig_PdfStrainI2}d for high packing fraction is magnified by the log scale, but the computation of the kurtosis evolution of $I_2$ does not show any clear change near $\phi^p_0$.} 

The distribution of the first principal invariant, $I_1$, has also been investigated. Observations similar to the ones made for $I_2$ have been concluded, so, \JB{by sake of clarity,} these results are not reported in this paper.

\begin{figure}[htb!]
	\begin{center}
		\includegraphics[width=0.35\textwidth]{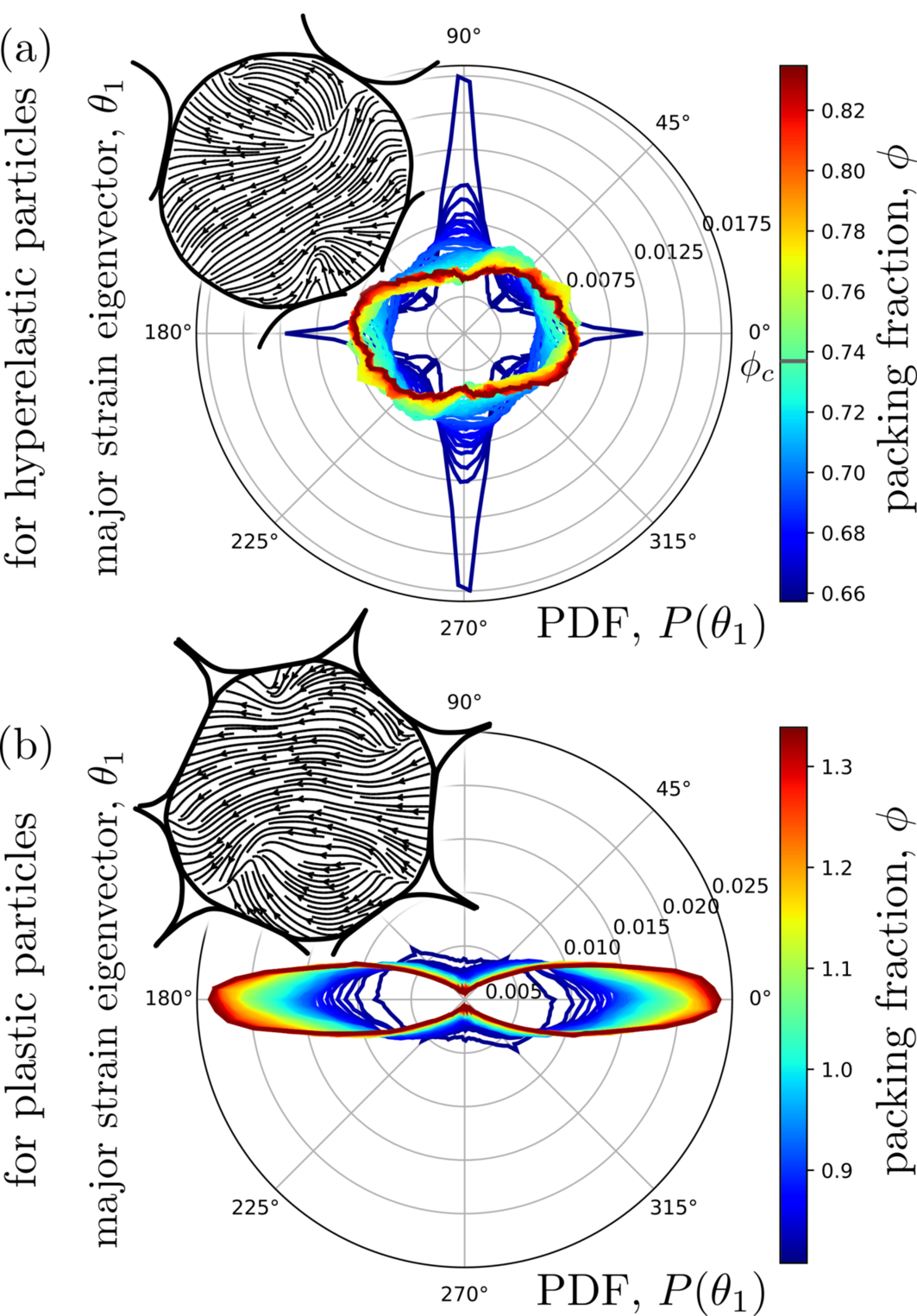}
	\end{center}
\caption{(Color online) Evolution of the probability density function (PDF) of the major strain eigenvector direction, $P(\theta_1)$, as a function of the packing fraction, $\phi$, for hyperelastic (a) and plastic grains (b). Results are presented for a single experiment for each material. \JB{Insets: major strain eigenvector field for large hyperelastic and plastic particles respectively. The measure is made for a packing fraction close to $\phi_0$.}}
\label{Fig_PdfMajorStrain}
\end{figure}

Figure~\ref{Fig_PdfMajorStrain} shows how the PDF of the \JB{first} eigenvector direction $P(\theta_1)$ evolves during the compression process for hyperelastic (a) and plastic (b) particles. In both cases when the system is fewly compressed no preferred direction is observed and the PDF is isotropic or mainly directed along the $x$ and $y$ axis. In the hyperelastic case, for higher packing fraction values, $P(\theta_1)$ tends to elongate in the direction normal to the compression direction as \JB{shown} in fig.~\ref{Fig_PdfMajorStrain}a \JB{and its inset where the eigenvector field is given for a particle at packing fraction close to $\phi_0$}. As shown in fig.~\ref{Fig_PdfMajorStrain}b this tendency is much stronger for plastic particles since $P(\theta_1)$ forms two lobes aligned along the horizontal axis. This mean that the intensity of the strain is larger in the normal direction to the compression which is consistent with the fact that particles are mainly \JB{squeezed} in this direction. We note that a similar study has been carried out \JB{for the second eigenvector}. Similar graphs rotated by $90^{\circ}$ are observed. These results are not presented in this paper.

\subsection{Local strain evolution} \label{sec_locX}

\begin{figure}[htb!]
	\begin{center}
		\includegraphics[width=0.45\textwidth]{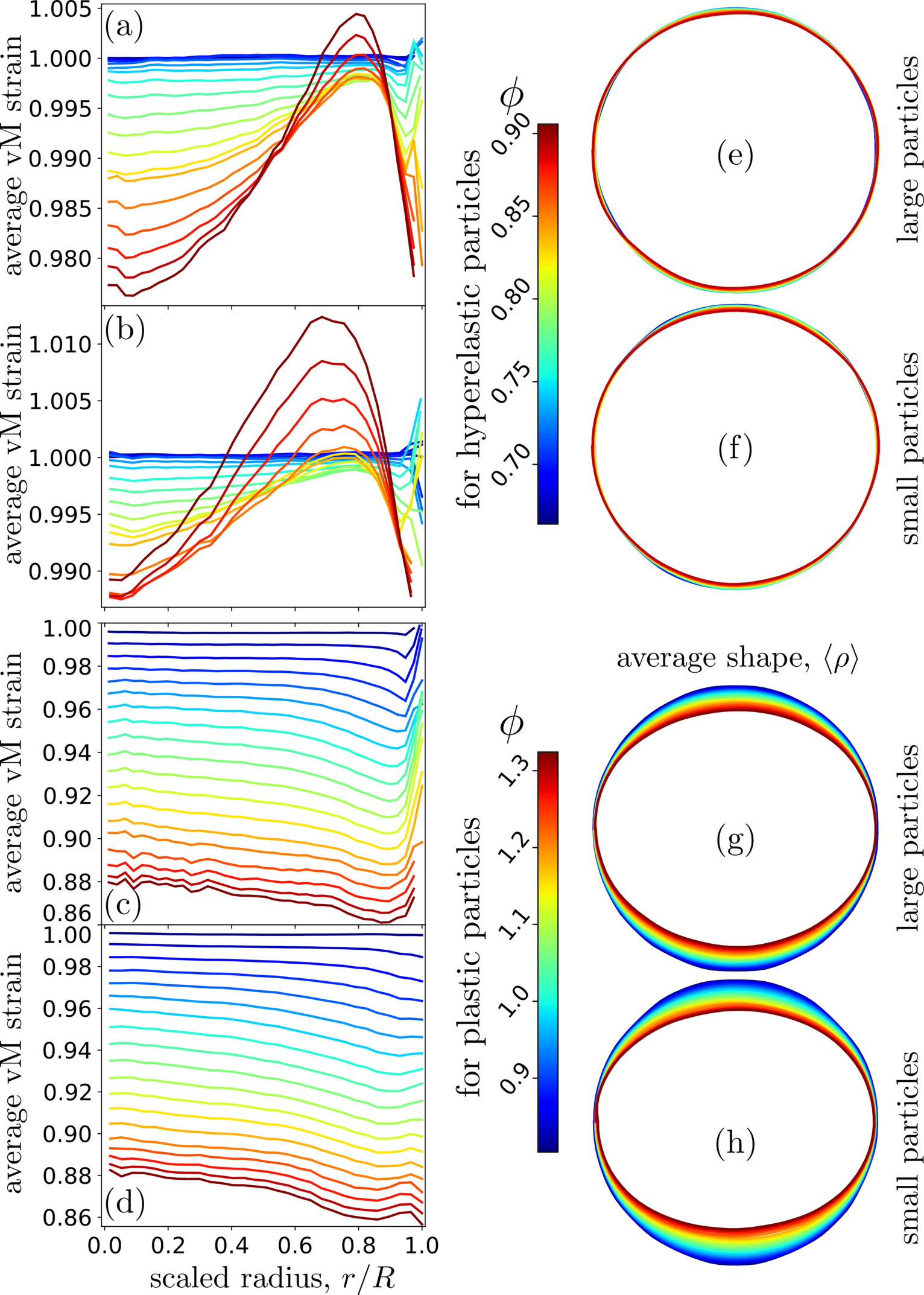}
	\end{center}
\caption{(Color online) left: Von Mises strain as a function of the scaled radius, $r/R$. Strain is \JB{averaged} radially for each particles over three experiments. The evolution of the curves is shown \JB{by varying the} packing fraction, $\phi$. Results are shown in a and b for hyperelastic grains and in c and d for plastic ones. Results are shown in a and c for large grains and in b and d for small ones. right: Evolution of the average grain shape as a function of the packing fraction, $\phi$. Results are shown in \JB{e and f} for hyperelastic grains and in \JB{g and h} for plastic ones. Results are shown in \JB{e and g} for large grains and in \JB{f and h} for small ones. Results are \JB{averaged} over three runs for each material.}
\label{Fig_AvgShape}
\end{figure}

In fig.~\ref{Fig_AvgShape}a to d, for each particle for different packing fractions, the von Mises strain field $\mathcal{C}$ has been averaged radially to get $\langle \mathcal{C} \rangle_{\theta}(r)$, where $(r,\theta)$ are polar coordinates attached to the particle. These different functions are then averaged over the particles -- distinguishing large and small ones -- over three different experiments for hyperelastic and plastic materials. As shown in fig.~\ref{Fig_AvgShape}a and b, in agreement with what is observed for a single particle \cite{vu2019_em} \JB{(see the inset of fig.\ref{Fig_PdfStrainVM}a)} , $\mathcal{C}$ is homogeneously equals to $1$ at the beginning of the compression and then rapidly decreases both in the center and very close to the edges of the particles. On the contrary, in between, few millimeters away from the edges, the von Mises strain increases and can even reach \JB{values} higher than $1$ in this area where the shear strain is dominant. We note \JB{that} this contrast between high and low $\mathcal{C}$ is higher for the smaller particles. 

\JB{As it was already suggested by the inset of fig.\ref{Fig_PdfStrainVM}d, fig.~\ref{Fig_AvgShape}c} and d show that the radial strain variation is completely different for plastic particles. For them, $\mathcal{C}$ decreases everywhere with $\phi$ and it decreases even faster in the area few millimeters away from the particle edges. This is due to the fact that plastic materials do not resist well shear strain and tend to plastically deform \JB{homogeneously in the whole particle}. Hence, the shear strain cannot reach high values compared to the hydrostatic strain; there is no area dominated by shear strain.

For each particle the average shape, $\langle \rho \rangle(\theta)$, for a given packing fraction has also been studied. For hyperelastic particles, fig.~\ref{Fig_AvgShape}e and f show that on average particles are almost not deformed except for high packing fractions where top and bottom edges move slightly inward. This comes from the fact that the material is incompressible and that, except for high $\phi$ values, the strain field is almost isotropic as emphasized by fig.~\ref{Fig_PdfMajorStrain}a. As shown in fig.~\ref{Fig_AvgShape}g and h, on the contrary for plastic material, \JB{on average,} particles clearly and solely deformed along the compression axis. This is in agreement with the fact that the material is \JB{compressible with a very low Poisson ratio} and that the strain field is mainly oriented horizontally as shown in fig.~\ref{Fig_PdfMajorStrain}b. We also note that small particles are more deformed than larger ones.

\section{Concluding dicussions} \label{concl}

We have introduced a novel multi-scale experimental method for studying the mechanical interactions of granular matter down to the very local scale with an extremely good accuracy. This method is based on DIC and the use of a flatbed scanner to image granular media from the decimeter scale to the micrometer scale. This has been applied to the study of \JB{highly jammed soft granular materials} with two typical material behaviors, namely incompressible hyperelastic and compressible plastic materials. On top of the classical macroscopic measurements (stress, packing fraction, coordination, fraction of NR), this method has been proved to gives access to all the particle geometry observables (position, shape, asphericity, anisotropy, orientation), void geometry observables (area, asphericity, solidity), displacement and strain fields at the particle scale with a so far never reached accuracy. Our study of the strain field for these two sorts of materials shows that at the onset of rigidity as well as deep in the jammed state, $\bm{C}$ is a pertinent observable to consider in order to characterize the granular matter behavior from the microscopic scale. \JB{More precisely, the average value and standard deviation of ($i$) the von Mises strain $\mathcal{C}$, of ($ii$) the eigenvalues and ($iii$) invariants of $\bm{C}$ shows changes of regime when crossing the jamming transition at $\phi_c$ and a second rigidity transition at $\phi_0$. This latter depends on the material nature and is believed to be caused by a rapid change in the competition between compression, dilation and shear at the material scale.} We believe it could replace the coordination \cite{hecke2009_jpcm} or the fraction of NR \cite{zhang2010_gm} as an observable that rule the system behavior.

\JB{We have also} observed that when compressing hyperlastic particles, the stress grows superlinearly with $\phi$ (see fig.~\ref{Fig_global}) which is not what is expected for Hookean particles \cite{majmudar2007_prl} nor for soap bubbles \cite{durian1995_prl} where a linear regime is observed. For particles made of plastic material, the behavior is again different since $\sigma$ grows sublinearly with $\phi$. This can be explained by the fact that the plastic modulus is lower than the elastic one \cite{vu2019_em} so the more the system is compressed, the more the material enters in the plastic regime and the less the system is stiff.

We have shown that on average, the specific contact lengths first increase linearly in the jammed regime and then slowly saturate for densely packed systems (see fig.~\ref{Fig_global_edge}) -- at least in the plastic case. The slopes corresponding with this linear regime greatly differs from one material to \JB{another. We believe this depends on the particle matter behavior and can constitutes a tool to probe the material properties of a system, in the biological case for instance} \cite{atia2017_arx,martin2004_dev}. We also note that contrary to what is observed from numerical simulations of \JB{deformable polygons}, deep in the \JB{jammed} state ($\phi>1$), the particle asphericities increases linearly with $\phi$.  

Considering the intergranular voids, we have also evidenced that they are very diverse in terms for size and shape. We have also shown that their total area is maximum for $\phi=\phi_c$, at the jamming point (see fig.~\ref{Fig_global_edge}). \JB{This means that most of the contact and the induced voids are formed at the jamming point and that, then, their size only decreases due to the particle deformation.} In this study, even if our geometry is not ideal for such an analysis, the onset of rigidity has also been observed on many other observables: ($i$) the global stress increases from $\phi_c$, ($ii$) $Z$ \JB{roughly equals to} $4$ for $\phi=\phi_c$, ($iii$--$iv$) the particle average anisotropy--asphericity decreases--increases from $\phi_c$. This last point is in agreement with the fact that the jamming transition can be observed from the particle shape as already stated in previous numerical studies \cite{bi2015_nat,bi2016_prx,boromand2018_prl,rodney2011_msmse}. At the local scale the different following observables also undergo measurable crossovers around the jamming point: $\sigma_{\mathcal{C}}$, $m_{\lambda_1}$, $\sigma_{\lambda_1}$, $m_{I_2}$ and $\sigma_{I_2}$.

Beyond the change in mechanical behavior observed at the jamming transition, another crossover in the material rigidity is evidenced deep in the jammed regime. At the global scale, for plastic systems, for $\phi=\phi^p_0\approx0.99$, $\mathcal{L}(\phi)$ exits a linear regime while $\mathcal{A}^p_s(\phi)$ enters a linear regime (see fig.\ref{Fig_global_edge}). At the local scale, for the same packing fraction ($i$) the standard deviations of the von Mises strain and of $\lambda_1$ undergo a crossover, ($ii$) the mean value of $\lambda_1$ reaches a maximum and ($iii$) $\sigma_{I_2}(\phi)$ exits a linear regime. For hyperelastic particles, for $\phi=\phi^h_0\approx0.795$, $m_{\mathcal{C}}$ reaches a minimum value and $\sigma_{\mathcal{C}}(\phi)$ enters a linear regime. This crossover deep in the jammed regime around $\phi_0$ evidences a microscopic change in the granular matter behavior that has never been observed before. We believe it is due to the material properties, the compression history and the particle geometry. \JB{Indeed from the observations of the local fields, we have suggested that rapid changes in the competition between compression, dilation and shearing at the local scale can induce these crossovers. The variation of these quantities deeply depends on the material properties and particle geometries so should $\phi_0$ do.}         

We have observed that on average hyperelastic particles are deformed more isotropically than the plastic ones \JB{that} are mainly deformed along the compression direction (see fig.~\ref{Fig_PdfMajorStrain} and \ref{Fig_AvgShape}). We believe this is due to the fact that hyperelastic particles that are more rigid and have more the ability to rearrange to minimize the stress inside the packing while plastic particles deform plastically to obtain the same result. Dealing with particle shape, we also note that the dense packing shown in fig.~\ref{Fig_expe}f is reminiscent of what is observed in biology for epithelial cells for example \cite{martin2004_dev,atia2017_arx} or in soap bubble experiments \cite{bolton1990_prl,katgert2010_epl}.      

\JB{We believe this work could become a reference point to build up a, currently lacking, theory of the mechanical properties of material made of soft particles. It could also be used to benchmark numerical models simulating the compression of these sorts of materials.} Beyond the study of highly strained soft discs packing we believe the experimental \JB{tool} introduced in this paper offers an opportunity to catch experimentally the behavior of breakable, polydisperse (in size, shape and material) and inflating granular matter under different sorts of loadings. It could even be adapted to the study of porous materials.

\section{Acknowledgements} \label{ackn}

Gille Camp, St\'{e}phan Devic  and R\'{e}my Mozul are greatly thanked for their technical support. Mathieu Renouf, \'{E}milien Az\'{e}ma and Vincent Huon are thanked for fruitful discussions.

\bibliographystyle{unsrtnat}
\bibliography{biblio}

\end{document}